\documentclass[
aps,%
11pt,%
final,%
notitlepage,%
oneside,%
onecolumn,%
twocolumn,%
nobibnotes,%
nofootinbib,%
superscriptaddress,%
noshowpacs,%
centertags]%
{revtex4}

\begin{document}



\title{Low Density Structures in the Local Universe.\\ I. Diffuse
Agglomerates of Galaxies}

\author{\firstname{I.~D.}~\surname{Karachentsev}}
\affiliation{\saoname}

\author{\firstname{V.~E.}~\surname{Karachentseva}}
\affiliation{Main Astronomical Observatory, National Academy of
Sciences, Kiev, 03680 Ukraine}

\author{\firstname{O.~V.}~\surname{Melnyk}}
\affiliation{Astronomical Observatory, Taras Shevchenko National
University of Kiev, 04053 Ukraine} \affiliation{Institut
d'Astrophysique et de Geophysique, Universit\'{e} de Li\`{e}ge,
Li\`{e}ge, B5C B4000 Belgium}

\author{\firstname{A.~A.}~\surname{Elyiv}}
\affiliation{Main Astronomical Observatory, National Academy of
Sciences, Kiev, 03680 Ukraine} \affiliation{Institut
d'Astrophysique et de Geophysique, Universit\'{e} de Li\`{e}ge,
Li\`{e}ge, B5C B4000 Belgium}

\author{\firstname{D.~I.}~\surname{Makarov}}
\affiliation{\saoname}

\received{August 30, 2012}  \revised{September 7, 2012}

\begin{abstract}
This paper is the first of a series considering the properties of
distribution of nearby galaxies in the low density regions. Among
7596 galaxies with radial velocities  $V_{\rm LG}<3500$~km/s,
absolute magnitudes  $M_K<-18\fm4$, and Galactic latitudes $|b|
>15^{\circ}$ there are 3168 field galaxies (i.e. 42\%) that do not belong to pairs,
groups or clusters in the Local universe. Applying to this sample
the percolation method with a radius of $r_0=2.8$~Mpc, we found
226 diffuse agglomerates with $n\geq4$ number of members. The
structures of eight most populated objects among them ($n\geq25$)
are discussed. These non-virialized agglomerates are characterized
by a median dispersion of radial velocities of about $170$~km/s,
the linear size of around $6$~Mpc, integral $K$-band luminosity of
$3\times10^{11}~L_{\odot}$, and a formal virial-mass-to-luminosity
ratio of about $700~M_{\odot}/L_{\odot}$. The mean density
contrast for the considered agglomerates is only $\langle\Delta
n/\overline{n}\rangle\sim 5$, and their crossing time is about
\mbox{$30$--$40$}~Gyr.
\end{abstract}

\maketitle

\section{INTRODUCTION}

Recent photometric and spectral massive sky surveys:
2MASS~\cite{Jar2000:Karachentsev_n},
SDSS~\cite{Abaz2009:Karachentsev_n},
2MRS~\cite{Huch2012:Karachentsev_n},
6dF~\cite{Jon2009:Karachentsev_n} etc. demonstrate that the
principal elements of the large-scale structure of the Universe
are cosmic voids, embordered by  filaments and walls towards which
the galaxies are concentrated. In knots, at the intersection of
walls and filaments rich clusters of galaxies are emerging. The
glow of hot intergalactic gas in rich clusters makes them
outstanding objects in the \mbox{X-ray} maps of the sky. Numerical
simulations of the formation and evolution of the large-scale
structure sustain this pattern quite
well~\cite{Klyp2003:Karachentsev_n,Schaap2007:Karachentsev_n,Shand2011:Karachentsev_n,Klyp2011:Karachentsev_n}.
In the modern epoch ($z=0$)  virialized regions of groups and
clusters of galaxies, as well as the collapsing regions around
them concentrate about 74\% of all the galaxies, or about 90\% of
stellar mass. However, these dynamically ``advanced'' regions
occupy only about 5\% of the total
volume~\cite{Kar2012:Karachentsev_n}. The remaining 95\% of the
volume are occupied by about a quarter of all galaxies (or 10\% of
stellar mass) which are involved in the infinite cosmic expansion.

At present we are witnessing a somewhat paradoxical situation:
rich clusters of galaxies, their structure and evolution are
already investigated with enough detail, while the properties of
the principal elements of the cosmic volume (the voids, filaments
and walls) are so far only known in the most general outline. The
emphasis made on the study of ``tops'' of the large-scale
structure and the neglect of its ``roots'' renders the prevailing
approach quite asymmetric. One reason for this asymmetry is the
paucity of data available on the individual distances of galaxies.
The Extragalactic Distance Database ({\tt
http://edd.ifa.hawaii.edu}), created by Tully et
al.~\cite{Tul2009:Karachentsev_n} shows that the relative number
of galaxies with measured distances rapidly drops with increasing
distance, making up a small percentage as early as at $D\sim10$
Mpc.

\begin{figure*}[tb]
\setcaptionmargin{5mm} \onelinecaptionsfalse
\includegraphics[scale=0.85,bb=20 20 584 237]{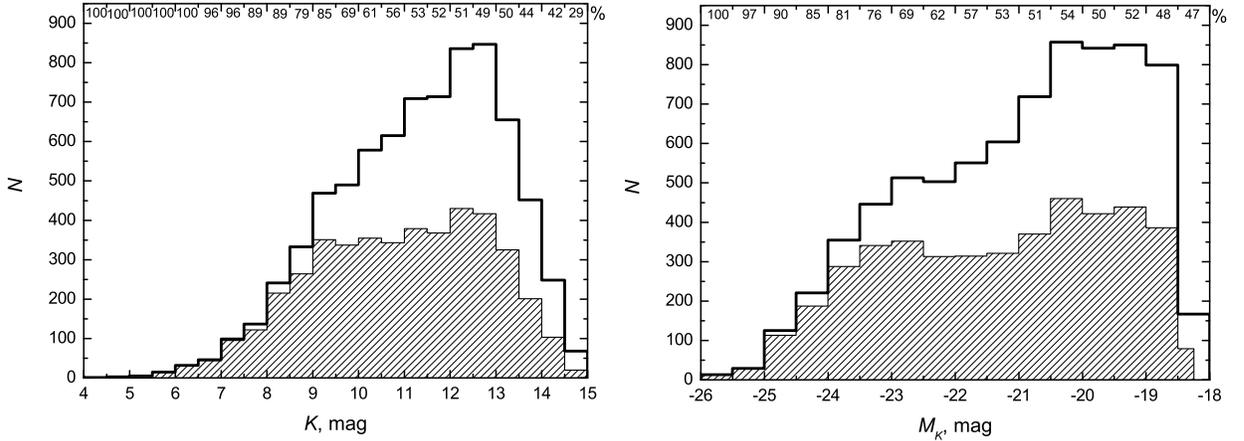}
\captionstyle{normal} \caption{The distribution of the number of
galaxies with radial velocities $V_{\rm LG}<3500$~km/s by apparent
($K$) and absolute $(M_K$) magnitudes in the $K$-band. Clustered
galaxies are hatched, their relative number in each interval is
shown in percentage at the top edge.}
\end{figure*}

\begin{figure}[tb]
\setcaptionmargin{5mm} \onelinecaptionsfalse
\includegraphics[scale=0.8]{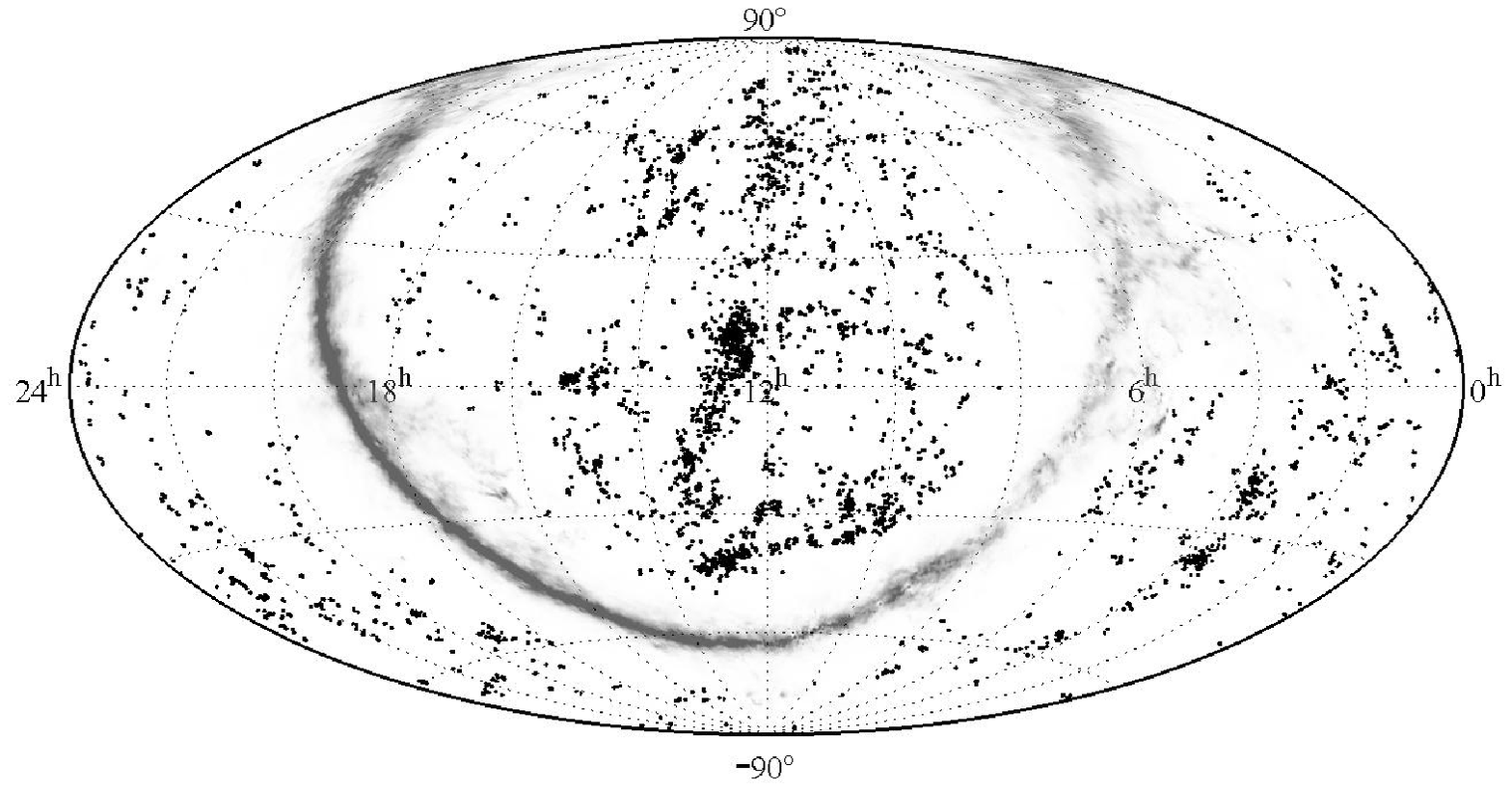}
\vspace{1mm}
\includegraphics[scale=0.8]{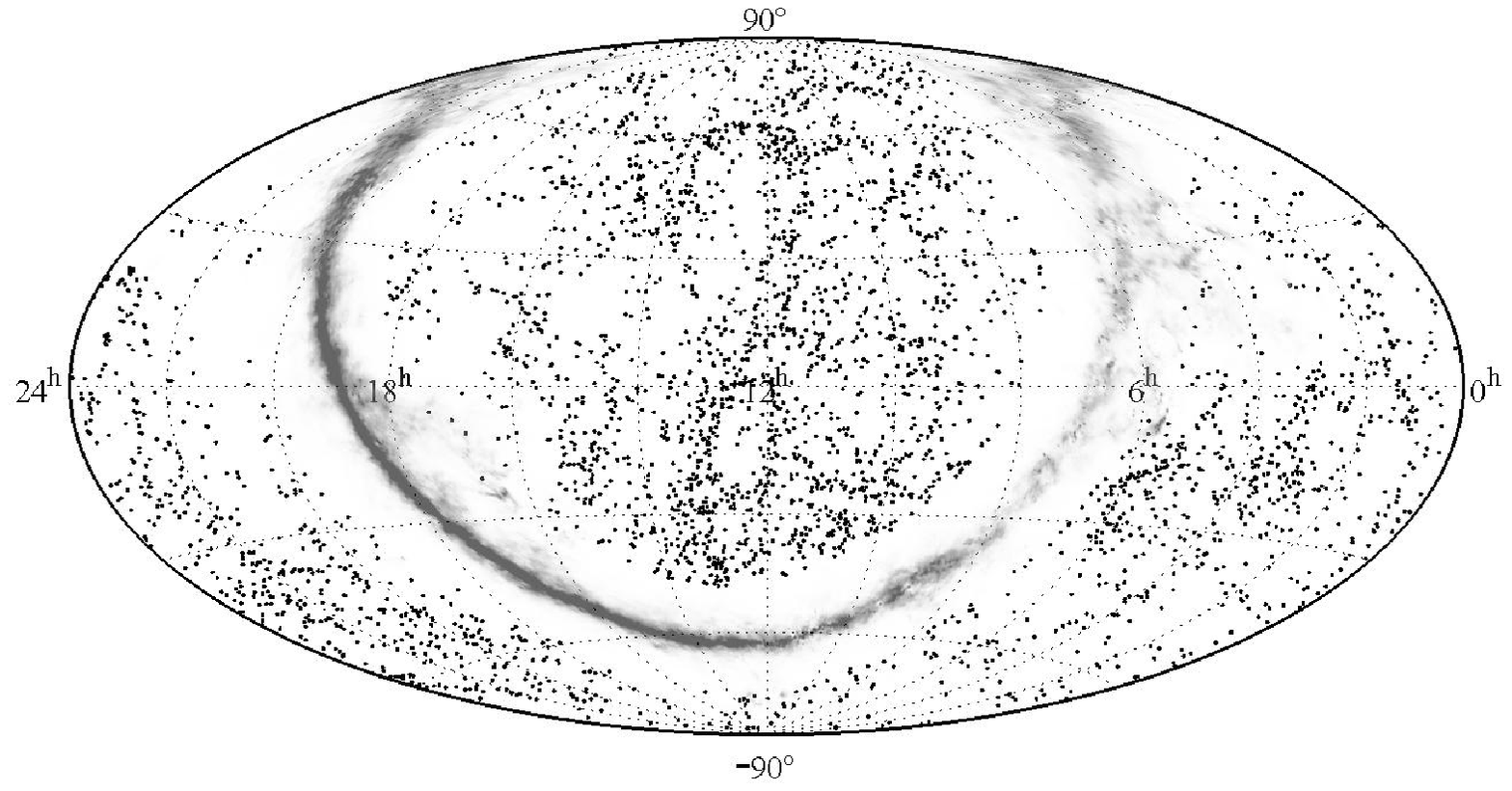}
\captionstyle{normal} \caption{The distribution of clustered (top)
and non-clustered (bottom) galaxies with \mbox{$V_{\rm LG}<3500$
km/s} in equatorial coordinates. The region of significant
galactic absorption with $A_g>2\fm0$ is described by a gray ragged
stripe.}
\end{figure}

To study the properties of the nearby part of the large-scale
structure, we compiled a sample of 10\,500 galaxies with radial
velocities of \mbox{$V_{\rm LG}<3500$}~km/s relative to the
centroid of the Local Group (LG), which covers the entire sky
except for low Galactic latitudes  $|b|<15^{\circ}$. The
morphological types, the data on the radial velocities and
apparent magnitudes were determined or refined for all the
galaxies of this sample. Using the new clustering algorithm, which
takes into account the galaxy differences by luminosity, there
were compiled the catalogs of 509
pairs~\cite{Kar2008:Karachentsev_n}, 168~triple
systems~\cite{Mak2009:Karachentsev_n} and 395 groups of
galaxies~\cite{Mak2011:Karachentsev_n}. In addition, a separate
catalog was devoted to 520 most isolated galaxies in this
volume~\cite{Kar2011:Karachentsev_n}.

Since the best indicator of the stellar mass of the galaxy is its
$K_s$-band luminosity, we adopted the apparent $K_s$-magnitudes of
galaxies from the 2MASS survey~\cite{Jar2000:Karachentsev_n}. In
their absence, the $K$-magnitudes of galaxies were determined from
the known \mbox{$B$-mag}ni\-tud\-es and the mean color indices
\mbox{$\langle B-K\rangle$} depending on the morphological type of
the galaxy. From the original sample of 10\,500 galaxies with
\mbox{$V_{\rm LG} = [0$--$3500]$}~km/s, we have eliminated the
objects fainter than $K=15\fm0$, and dwarf galaxies with absolute
magnitudes of \mbox{$M_K>-18\fm4$} given the Hubble constant of
$H_0=73$ km/s/Mpc. The latter condition provides that the galaxies
with luminosities brighter than the luminosity of the Small
Magellanic Cloud would be visible both nearby and at the far edge
 $(m-M=33\fm4$) of the considered volume. Having sacrificed 2906 dwarf
galaxies (28\% of the sample), we have considerably relaxed the
selection effect by distance, which was imposing unequal
conditions to the nearby and far volumes. We use the sample of
7596 galaxies corrected this way for the further analysis of
elements of the large-scale structure of the Local universe at
extremely low densities.

\section{THE FIELD GALAXIES AND CLUSTERED POPULATION}

The distribution of 7596 galaxies in our sample by the apparent
($K$) and absolute $(M_K)$ magnitudes is presented in the left and
right panels of Fig.~1. The maximum of the $N(K)$ distribution
falls on \linebreak $K\simeq12.5$, from what we can conclude that
in many galaxies of this volume with $K=12.5$--$15.0$ the
line-of-sight velocities have not yet been measured. The clustered
galaxies are marked by hatching in both panels. The clustering
algorithm applied assumes that the stellar mass of each galaxy is
determined by its $K$-luminosity, and the total mass of its dark
halo is $\kappa = 6$ times larger than the stellar mass. The
criterion of inclusion of galaxies in a pair or group was based on
two obvious considerations: 1)~the total energy of a hypothetic
system has to be negative, and 2) the members of a virtual system
have to be causally interrelated (their mutual separations have to
be within the ``zero-velocity sphere'', which separates a
potential group from the common Hubble expansion). The latter
condition is required as a complement to the former, since we do
not know the total (spatial) distances and velocities of galaxies.
In fact, our algorithm has an only arbitrary parameter $\kappa$,
assumed to be equal to 6 regardless of the luminosity of a galaxy
and its neighborhood. Note that the global ratio of the dark
matter density to the baryon density is
$\Omega_m/\Omega_b\simeq6$~\cite{Fuk2004:Karachentsev_n}.

The application of our algorithm resulted in the integration of
4428 galaxies out of 7596 into systems, i.e. the percentage of
clustered galaxies and members of the general ``field'' amounted
to 58:42. It proved to be slightly higher than for the entire
original sample, 52:48, from which the dwarf objects were not yet
eliminated. These figures suggest that normal galaxies tend to get
clustered more than their dwarf counterparts.

\begin{table*}
\setcaptionmargin{5mm} \onelinecaptionstrue
\captionstyle{nonumber} \caption{\centerline{Parameters of the
most populated agglomerates in the Local universe}}
\medskip
\begin{tabular}{|l|c|c|c|c|c|c|c|r|r|}
 \hline
Agglpmerate  & RA DEC& n& $<V_{LG}>$ & $\sigma_v$ & $<r_{12}>$& $L_K$ & $M_{VIR}$
& $M_{VIR}/L_K$& n(E,S0)\\
	   &       &  & km/s       &km/s        & Mpc       &$10^{11}L_{\odot}$
& $10^{14}M_{\odot}$ & & \\

 \hline
Leo--Virgo        &   $11.7^h +4^{\circ}$&   83 &  +1210 & 158&  8.7 & 3.4&  2.4&  700 & 9\\
Eridanus--Columba &   4.3 --36        &   69 &  +1080 & 273&  7.6 & 3.0&  6.2&
2050 & 13\\
Centaurus         &  13.3 --32        &   43 &  +2310 & 182&  6.5 & 4.3&  1.6&
360 & 4\\
Microscopium      &  21.0 --39        &   39 &  +2670 & 110&  6.2 & 3.0&  0.8&
270 &3\\
Crater--Corvus    &  11.9 --17        &   33 &  +1510 & 180&  4.8 & 1.7&  1.7&
1000 & 1\\
Libra--Hydra      &  15.1 --20        &   29 &  +2300 & 153&  6.2 & 2.3&  1.6&
690 &2\\
Virgo             &  12.4 +2          &   25 &  +2070 & 217&  4.7 & 2.0&  2.4&
1210 &1 \\
Tucana--Grus      &  22.5 --59        &   25 &  +3170 & 109&  4.1 & 5.1&  0.6&
110 &2\\
\hline
\end{tabular}
\end{table*}

The sky distribution of 4428 clustered and 3168  non-clustered
galaxies of the Local universe is presented in the top and bottom
panels of Fig.~2 in equatorial coordinates. The region of
significant galactic absorption along the Milky Way is shown by
the gray ragged stripe. As we can see, the members of systems of
different multiplicity are revealing a strong concentration
towards the equator of the Local Supercluster, centered in the
Virgo cluster $(12^{\rm h}30^{\rm m}$~$+12^{\circ}$). The
population of non-clustered galaxies practically does not show
this concentration. At the same time, the distribution of field
galaxies does not look quite uniformly random. In different
regions of the sky  low contrast structures are visible, the
presence of which is not related to the flocky galactic
extinction.

\begin{figure}[tb]
\setcaptionmargin{5mm} \onelinecaptionsfalse
\includegraphics[scale=0.85,bb=13 10 293 228]{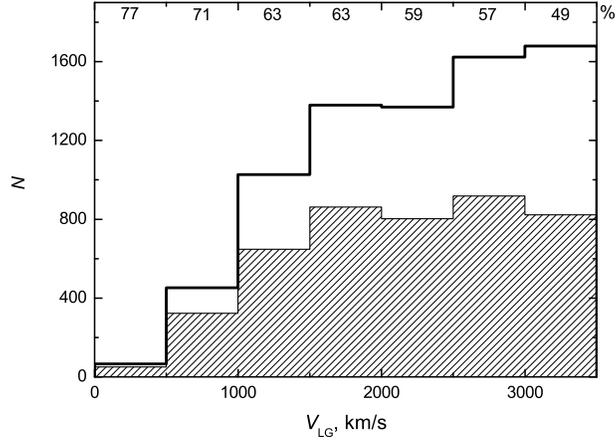}
\captionstyle{normal} \caption{ The distribution of galaxies in
the Local universe by radial velocities. Clustered galaxies are
highlighted by hatching, their relative number in each velocity
interval is marked in percentage at the top edge.}
\end{figure}

\begin{figure}[tb]
\begin{minipage}[t]{0.48\linewidth}
\setcaptionmargin{5mm}
\includegraphics[scale=0.85,bb=13 10 293 220]{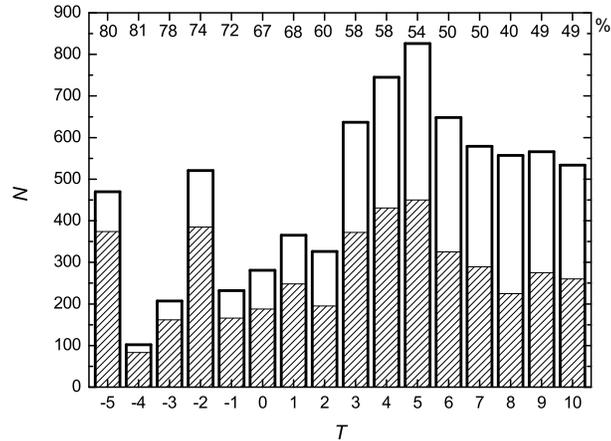}
\captionstyle{normal}\onelinecaptionsfalse \caption{Distribution
of the number of   galaxies by morphological type. Clustered
galaxies are hatched, their percentage in each type is indicated
at the top edge.}
\end{minipage}
\end{figure}

Figure~3 represents the line-of-sight velocity distribution of
galaxies in our sample in the LG centroid frame. Clustered
galaxies are marked by hatching. Their relative number next to the
ulterior boundary of the volume significantly drops due to the
shortage of galaxies with measured velocities among the distant
objects.

As we know, the galaxies of early morphological types (E, S0, Sa)
show a higher degree of clustering than the late-type galaxies.
The expected effect of morphological segregation is also evident
in our samples. Figure~4 shows the distribution of galaxies of our
volume by morphological types in the de Vaucouleurs scale. The
clustered galaxies are marked by hatching, their percentage for
each type is shown at the top edge. As we can see, the galaxies
with developed bulges  ($T<4$) are present among the clustered
galaxies in a much greater proportion than among the galaxies of
the field.
\begin{figure}[tb]
\begin{minipage}[t]{0.48\linewidth}
\setcaptionmargin{5mm}
\includegraphics[height=63mm,bb=13 10 283 228]{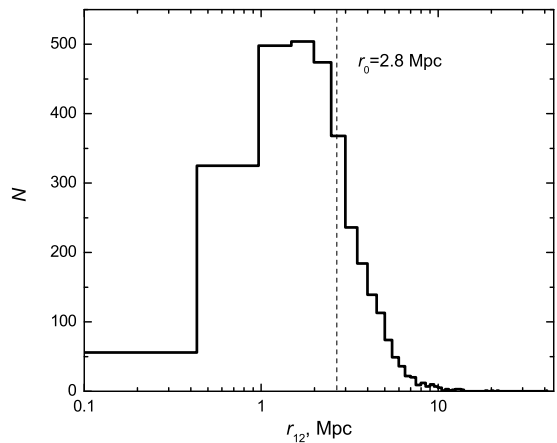}
\captionstyle{normal}\onelinecaptionsfalse \caption{Distribution
of non-clustered  galaxies by distance to the nearest neighbor.}
\end{minipage}
\end{figure}

\begin{figure*}[tb]
\begin{minipage}[t]{0.48\linewidth}
\setcaptionmargin{5mm}
\includegraphics[scale=0.55,bb=13 10 293 410]{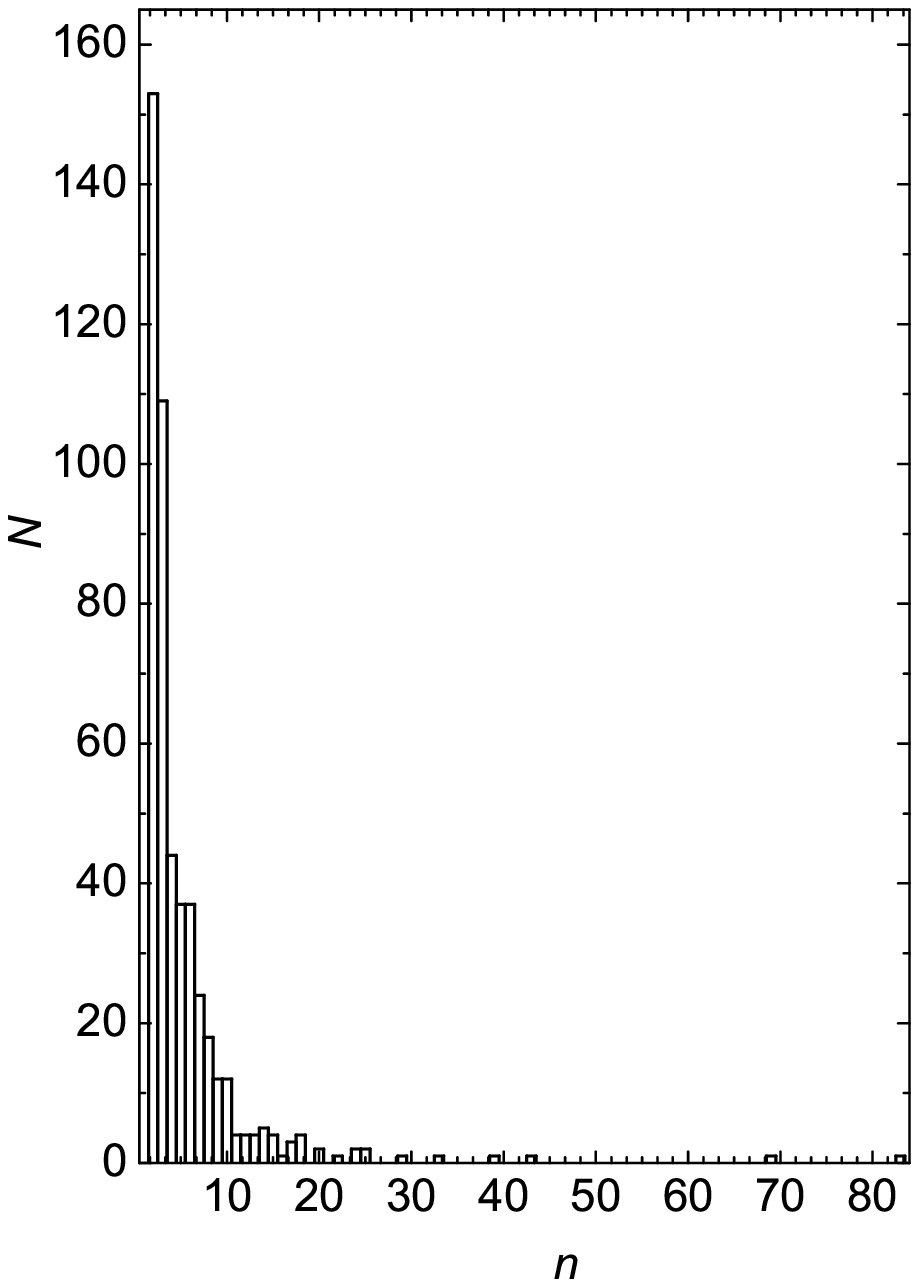}
\captionstyle{normal} \onelinecaptionsfalse \caption{The number of
low density agglomerates, identified by percolation, depending on
the number of galaxies in them.}
\end{minipage}
\begin{minipage}[t]{0.48\linewidth}
\setcaptionmargin{5mm}
\includegraphics[scale=0.55,bb=13 10 293 410]{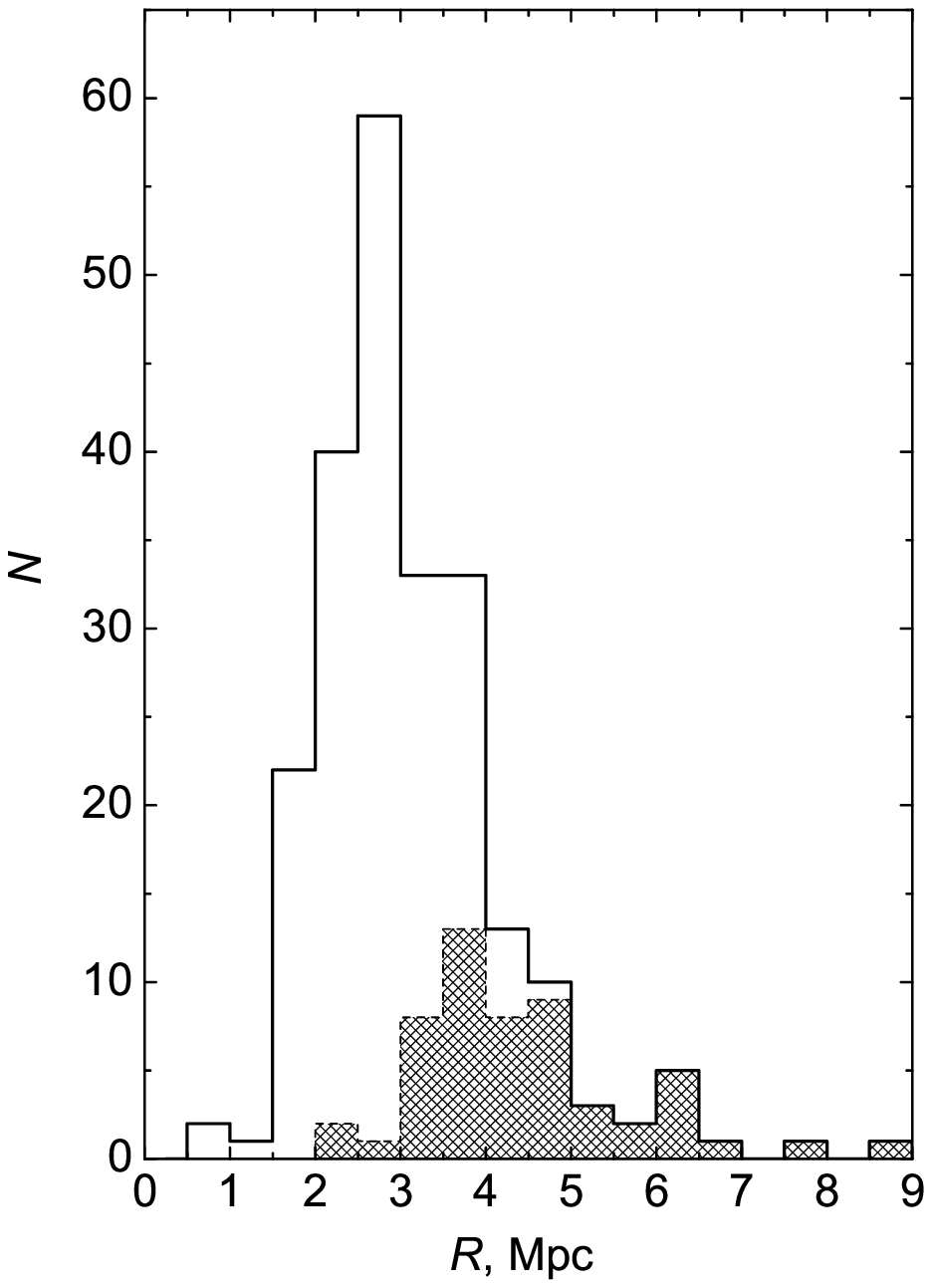}
\captionstyle{normal}\onelinecaptionsfalse \caption{Distribution
of the agglomerates with $n\geq 4$ number of members by the
average mutual separation of galaxies. The most populated
structures with $n\geq 10$ are shaded.}
\end{minipage}
\end{figure*}

\section{PERCOLATION AND LOW DENSITY GALAXY AGGLOMERATES}

To make headway in understanding the features of the distribution
of 3168 non-clustered galaxies, we tried to sort out any
non-random structures among them. This can be done by different
means. We used the simplest method of percolation. Estimating the
distances to galaxies by their radial velocities \mbox{$D=V_{\rm
LG}/H_0$} at \mbox{$H_0=73$~km/s/Mpc} and neglecting their
peculiar velocities, we have determined the spatial distance $r_
{12}$ to the nearest neighbor for each of them. The distribution
of the number of galaxies in increments of $r_{12}$ is shown in
Fig.~5 in the logarithmic scale. Two-thirds of the galaxies have
their nearest neighbor within $r_0=2.8$~Mpc.
We selected this value as the percolation radius via the ``friends
of friends'' (FoF) method ~\cite{Huch1982:Karachentsev_n}.

\begin{figure}[tb]
\setcaptionmargin{5mm} \onelinecaptionsfalse
\includegraphics[width=\columnwidth]{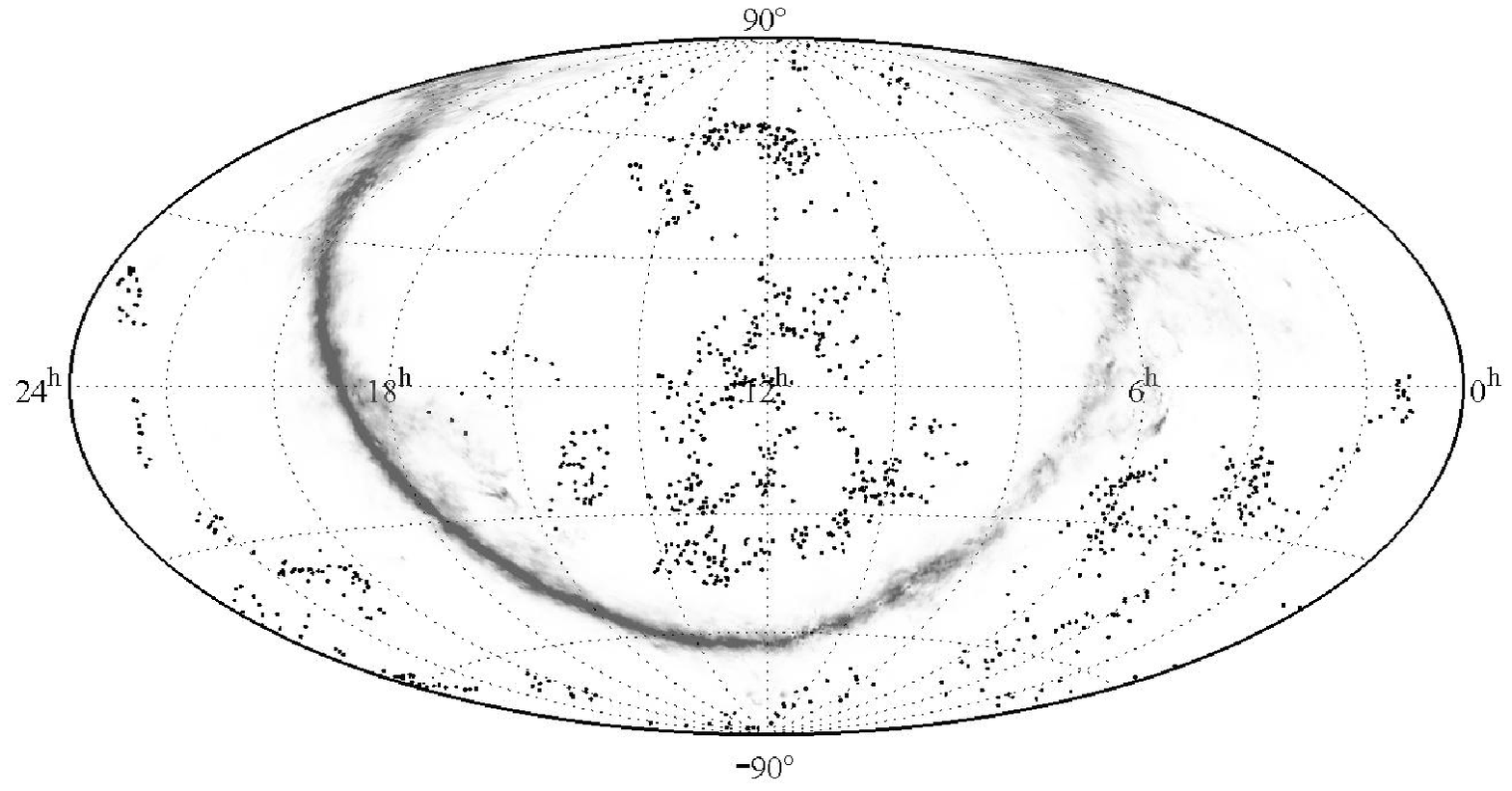}
\includegraphics[width=\columnwidth]{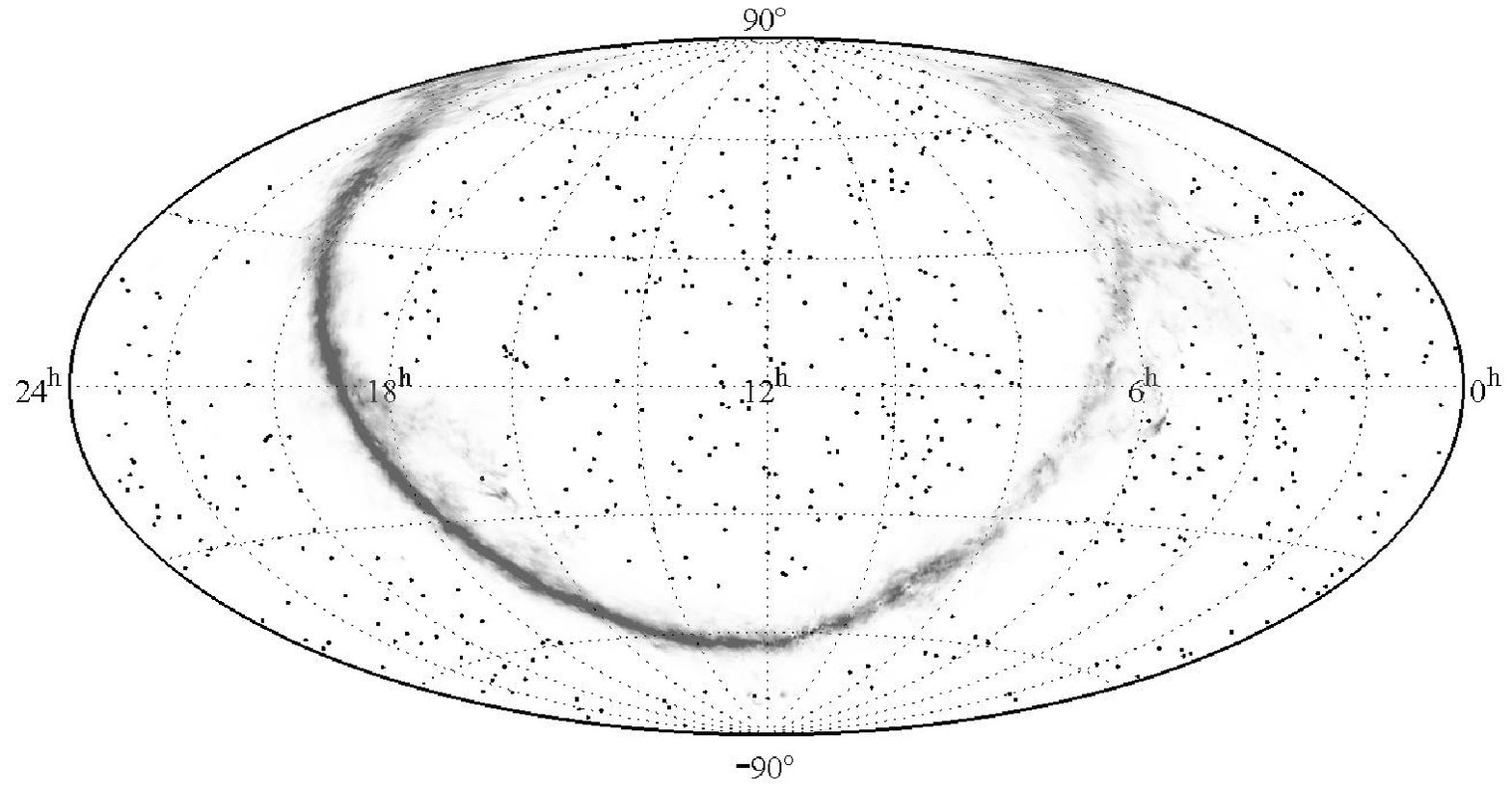}

\captionstyle{normal} \caption{The sky distribution of galaxies,
belonging to the agglomerates with $n\geq 10$ members (top panel)
and galaxies, not subjected to percolation (bottom panel), in
equatorial coordinates.}
\end{figure}

Combining the galaxies with mutual separations of   $r_{12}<2.8$
Mpc into the agglomerates of different populations $n$, we
obtained the following result: the number of non-percolated, i.e.
very isolated galaxies was found to be 543. The remaining galaxies
have grouped into agglomerates with the number of members from 2
to 84. The distribution of the number of such structures by the
number of members is shown in Fig.~6. It is easy to see that
compared with the Poisson distribution this one has a long tail,
the presence of which indicates that the unification of galaxies
in large associations is not random.

Figure~7  shows the distribution of 226 agglomerations with
$n\geq4$ members, as well as 54 most populated structures with
$n\geq10$ by the average mutual separation of their members. The
medians of these distributions are 2.9 Mpc and 4.2 Mpc,
respectively, i.e. the linear dimensions of these structures are
comparable to the virial radius of rich clusters of galaxies.

The sky  distribution of 989 galaxies belonging to the
agglomerates with $n\geq10$  members is demonstrated in the upper
panel of Fig.~8. For a comparison,   the bottom panel displays a
similar distribution of 543 single galaxies, not subjected to
percolation ($n=1$). The characters of these distributions are
strikingly different, once again suggesting that the low density
regions are hosting some non-virialized extended structures, which
comprise a significant number of galaxies.

\begin{figure*}[tb]
\setcaptionmargin{5mm} \onelinecaptionsfalse
\vspace{3mm} \centerline{ \hbox{
\includegraphics[scale=0.70]{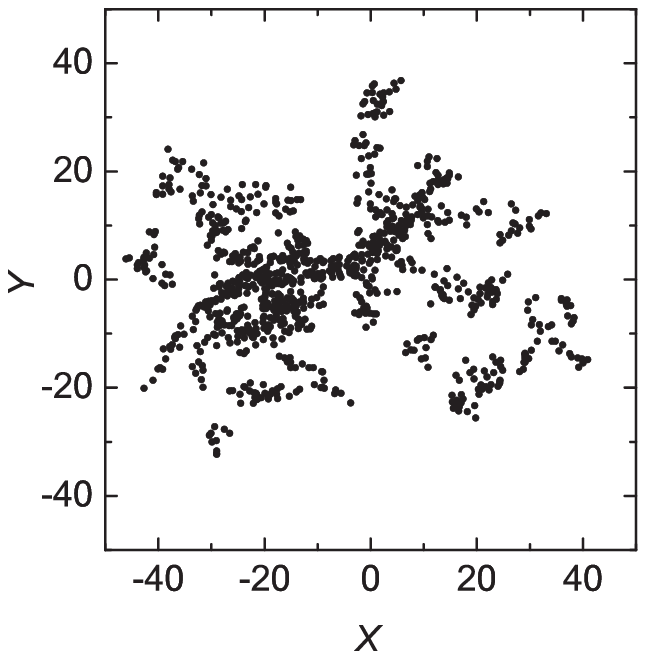}
\hspace{5mm}
\includegraphics[scale=0.70]{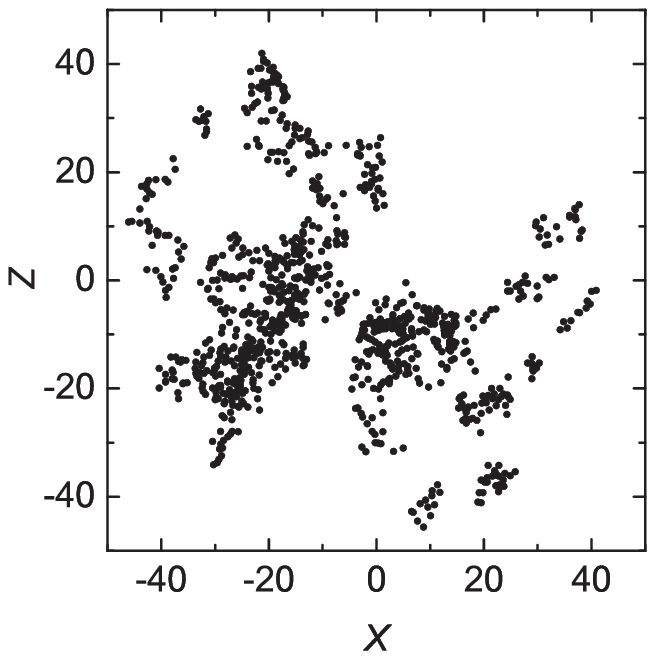}
\hspace{5mm}
\includegraphics[scale=0.70]{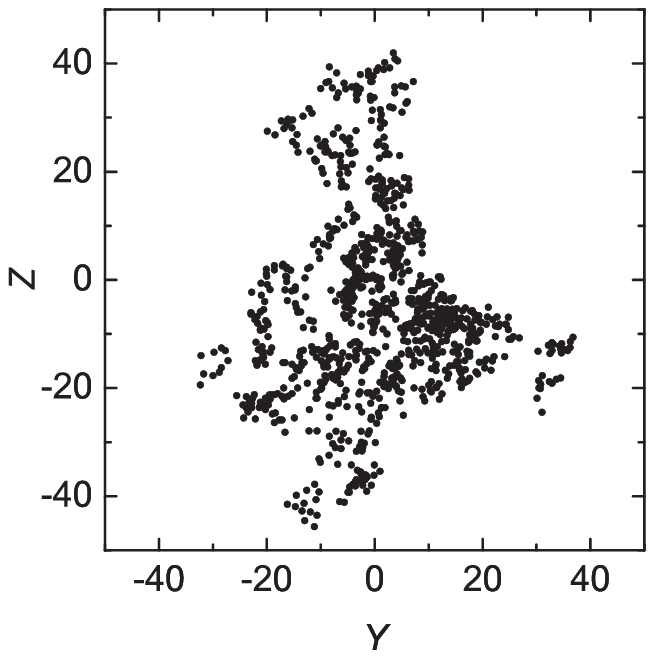}
}} \captionstyle{normal} \caption{The spatial distribution of
members of the low density agglomerates with $n\geq10$ presented
in three planes of equatorial Cartesian coordinates (in Mpc).}
\end{figure*}

The flocky nature of the distribution of galaxies belonging to the
agglomerates with populations of $n\geq10$ is also clearly visible
in Fig.~9, where we used the Cartesian equatorial coordinates. The
spottiness of these projected distributions is  partly caused by
the presence of the extinction region in the Milky Way. However,
the filamentary structure of a great many agglomerates can not be
caused by the effect of galactic extinction only.

\section{THE MOST POPULATED GALAXY AGGLOMERATES}

A summary of eight diffuse agglomerates in the Local universe with
the  $n\geq25$ number of galaxies is presented in the table. The
columns of the table include: (1) the names of the constellations,
where the agglomerate is located, (2) the equatorial coordinates
of the centroid, (3) the number of members with measured radial
velocities, (4) the mean line-of-sight velocity relative to the
Local Group, (5) the line-of-sight velocity dispersion, (6) the
mean spatial separation between the agglomerate members, (7)
integral $K$-band luminosity (i.e. the total stellar mass), (8)
the formal value of virial mass, (9) the formal virial
mass-to-$K$-luminosity ratio (or the  ratio of dark-to-luminous
matter), (10) the number of the agglomerate members of E and S0
morphological types. As we can see, the relative number of
early-type galaxies in these structures is only about 10\%.

\begin{figure*}[p]
\setcaptionmargin{5mm} \onelinecaptionsfalse \centerline{\vbox{
\hbox{
\includegraphics[scale=0.73]{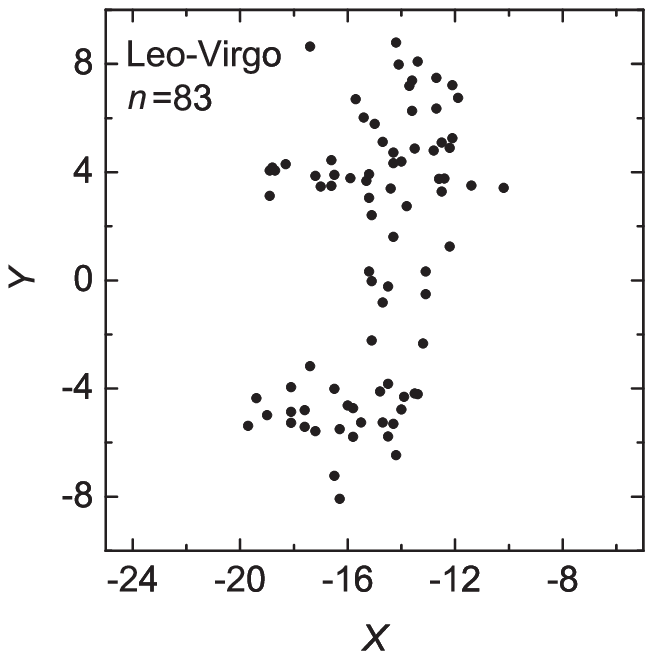}
\hspace{6mm}
\includegraphics[scale=0.73]{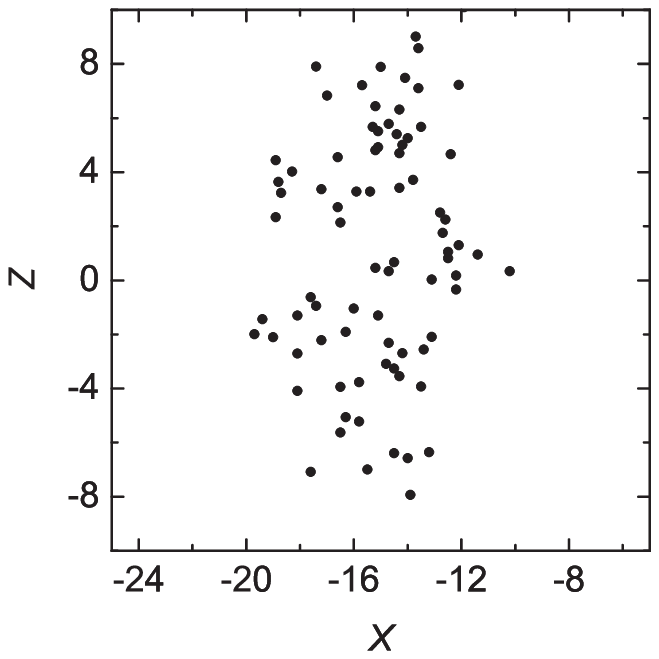}
\hspace{6mm}
\includegraphics[scale=0.73]{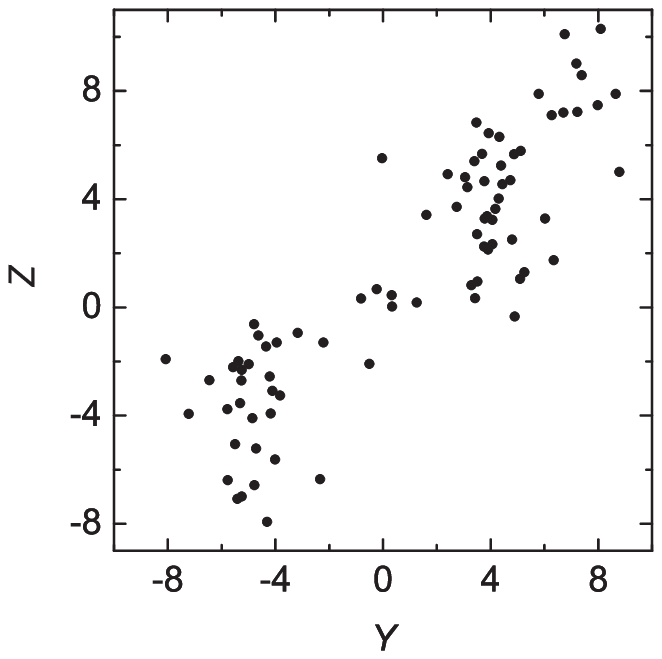}
} \vspace{5mm} \hbox{
\includegraphics[scale=0.73]{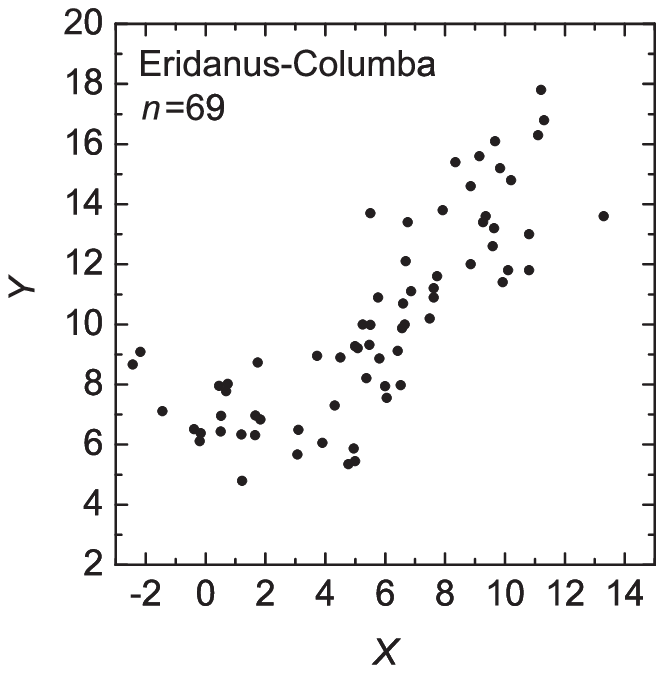}
\hspace{5mm}
\includegraphics[scale=0.73]{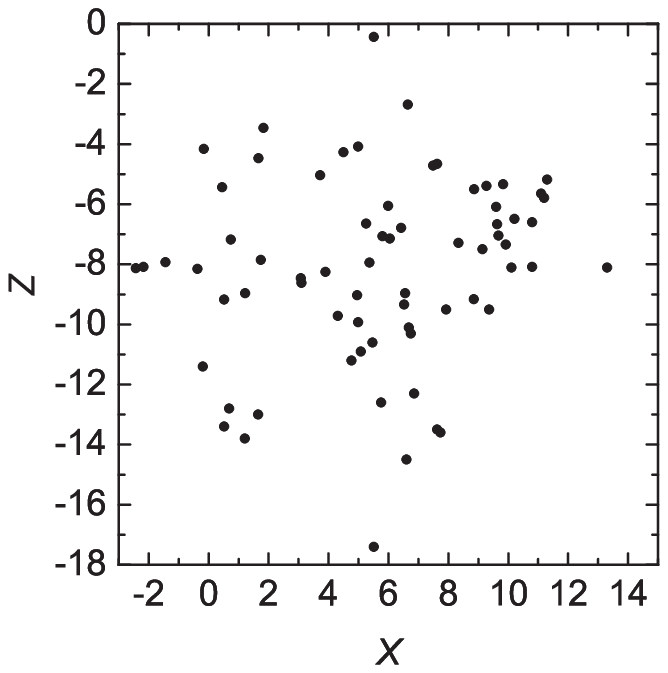}
\hspace{5mm}
\includegraphics[scale=0.73]{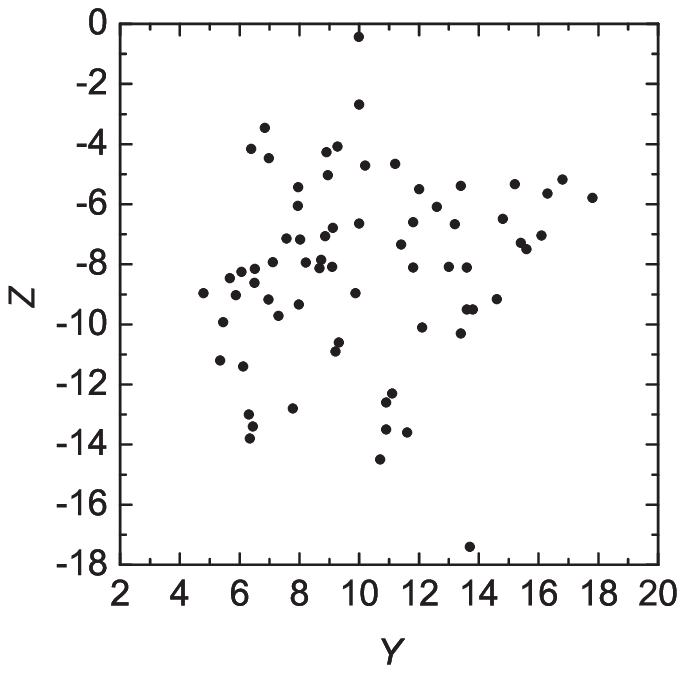}
} \vspace{5mm} \hbox{
\includegraphics[scale=0.73]{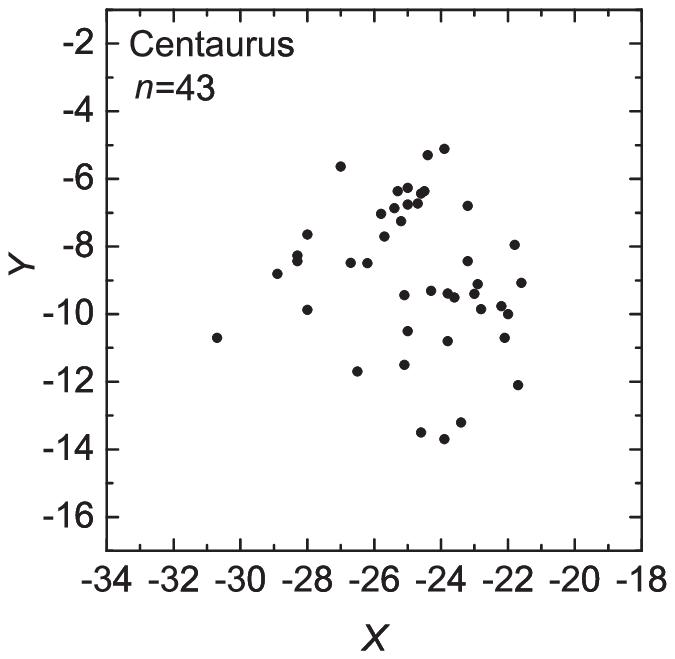}
\hspace{4mm}
\includegraphics[scale=0.73]{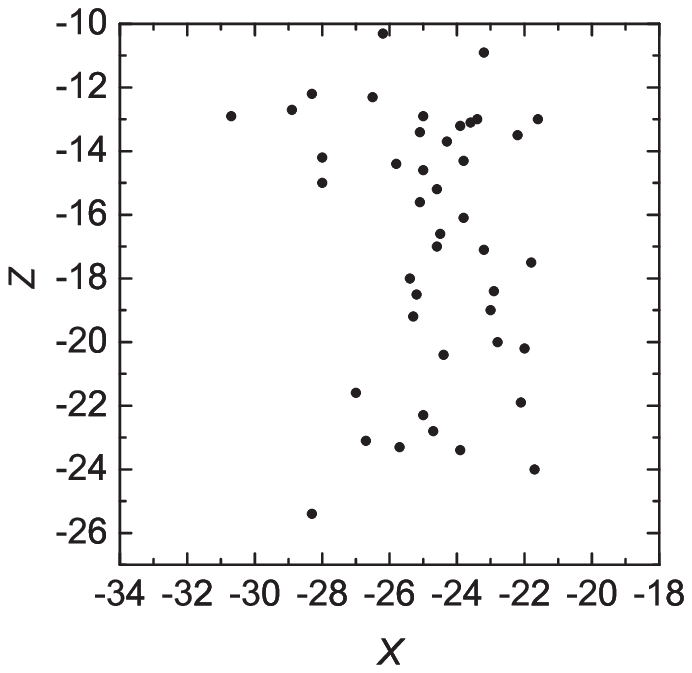}
\hspace{4mm}
\includegraphics[scale=0.73]{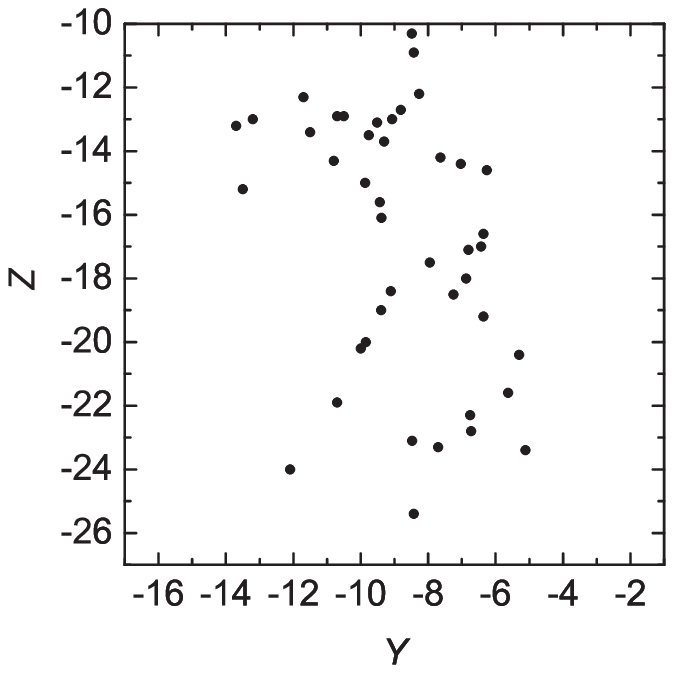}
} \vspace{5mm} \hbox{
\includegraphics[scale=0.73]{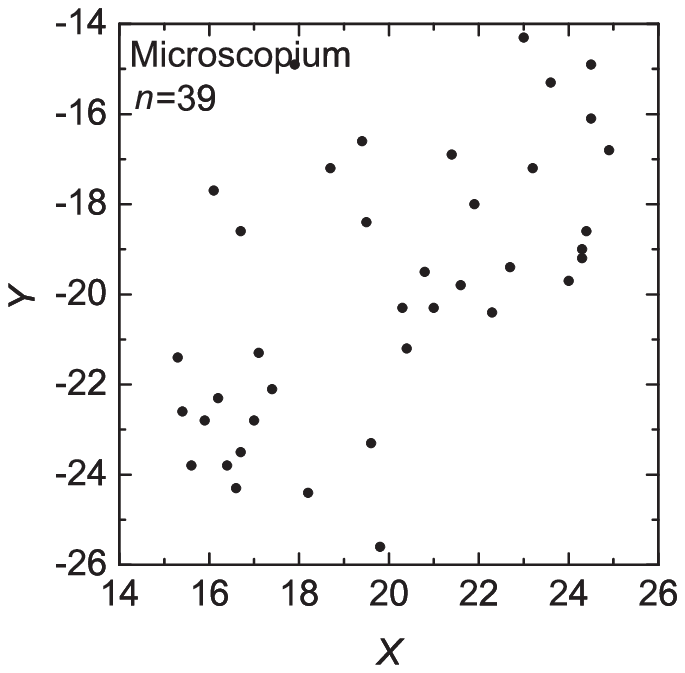}
\hspace{4mm}
\includegraphics[scale=0.73]{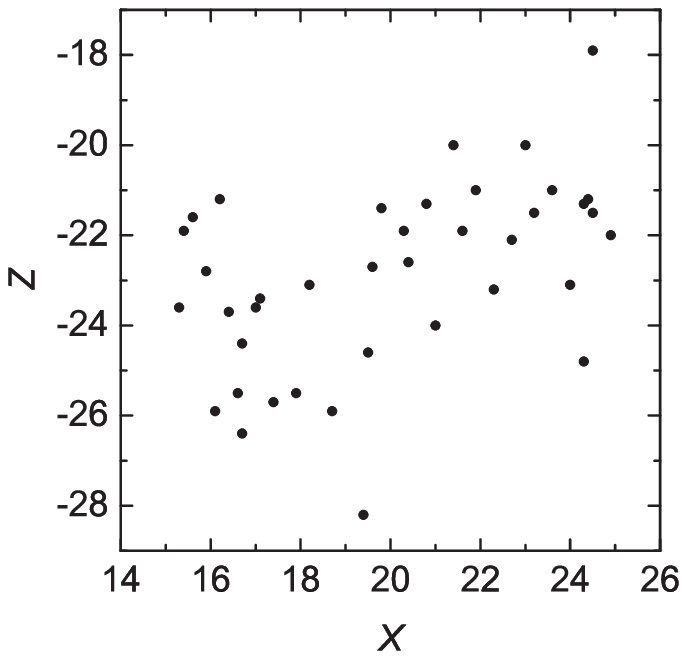}
\hspace{4mm}
\includegraphics[scale=0.73]{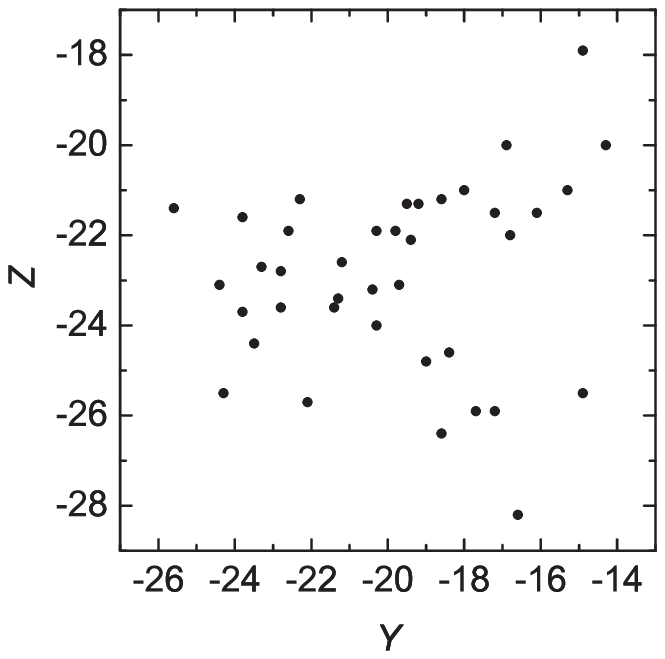}
}}}
\captionstyle{normal} \caption{The structure of eight most
populated low-density agglomerates in the projections of
equatorial Cartesian coordinates (in Mpc). The first part,
continued in Fig.~11.}
\end{figure*}

\begin{figure*}[p]
\setcaptionmargin{5mm} \onelinecaptionsfalse \centerline{\vbox{
\hbox{
\includegraphics[scale=0.73]{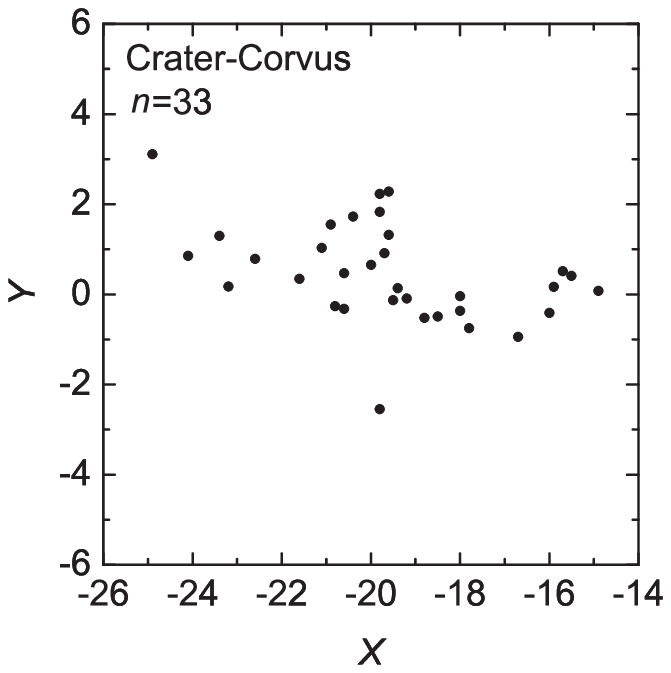}
\hspace{4mm}
\includegraphics[scale=0.73]{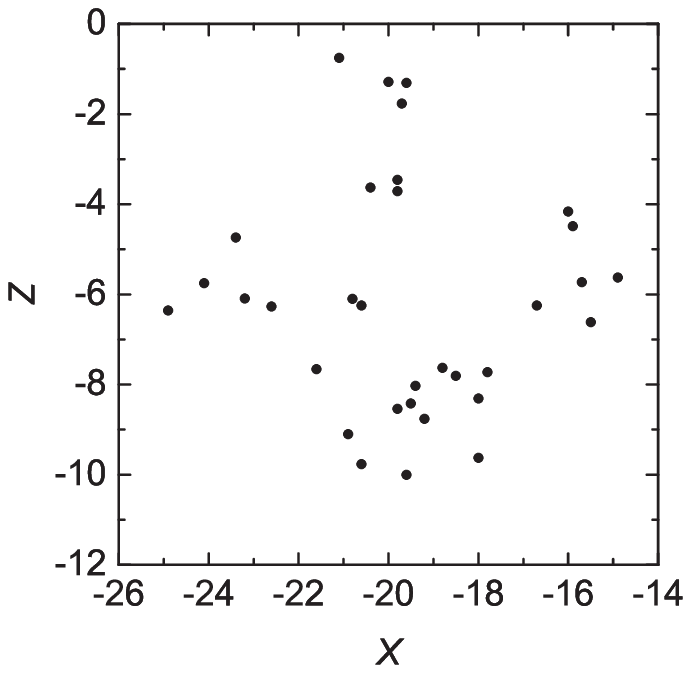}
\hspace{4mm}
\includegraphics[scale=0.73]{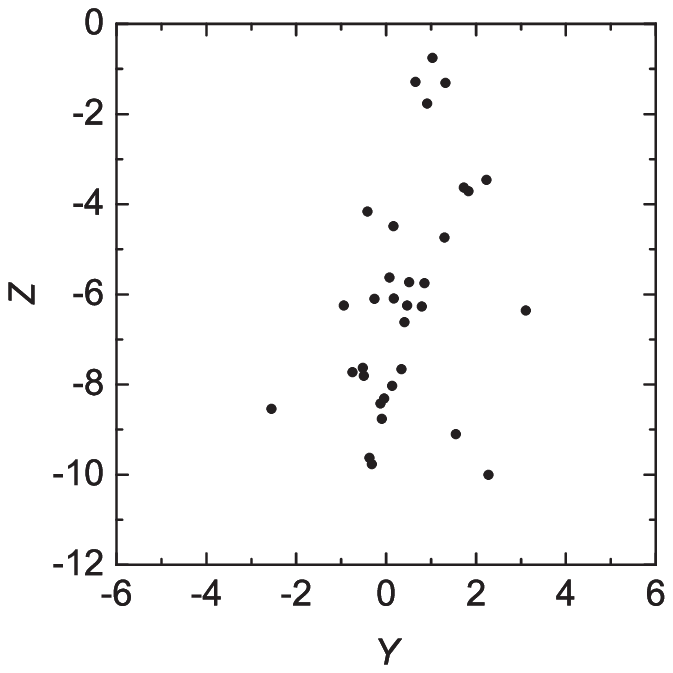}
} \vspace{5mm} \hbox{
\includegraphics[scale=0.73]{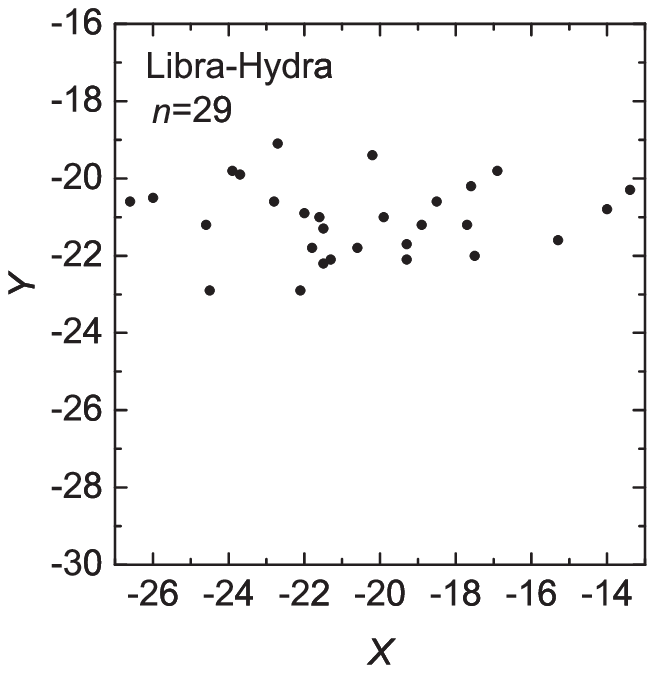}
\hspace{6mm}
\includegraphics[scale=0.73]{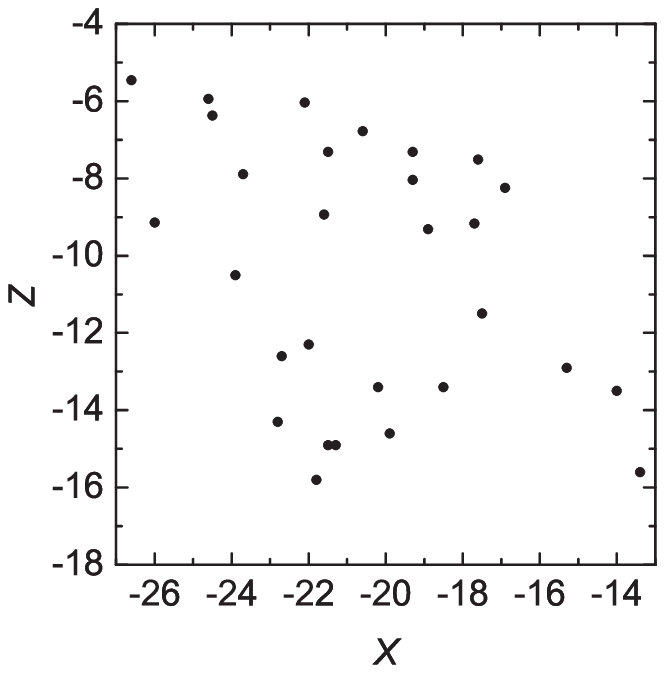}
\hspace{6mm}
\includegraphics[scale=0.73]{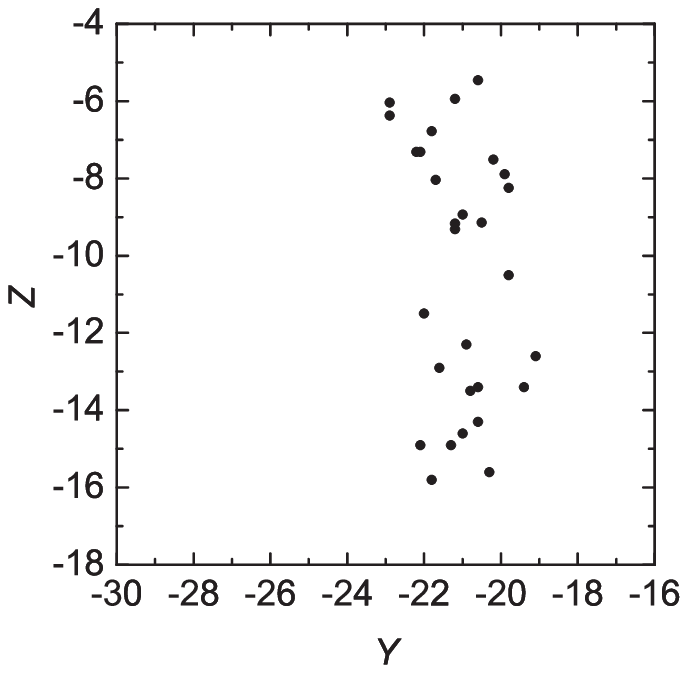}
} \vspace{5mm} \hbox{
\includegraphics[scale=0.73]{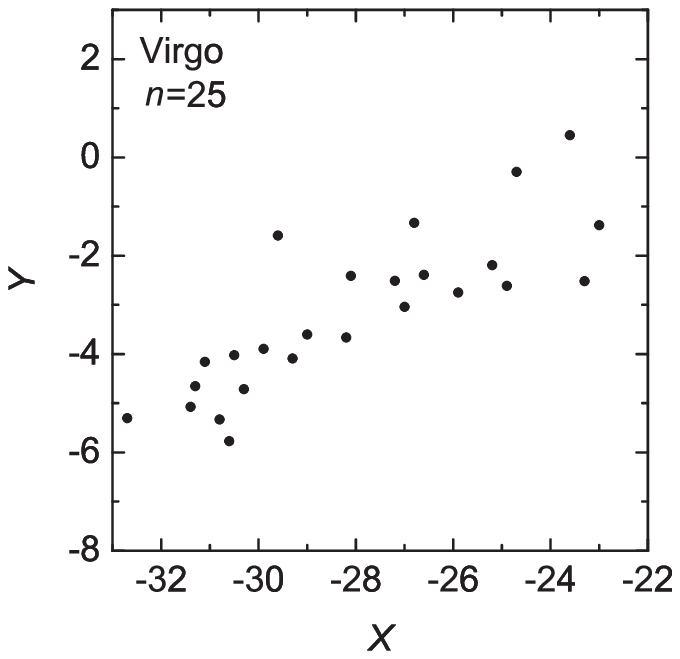}
\hspace{5mm}
\includegraphics[scale=0.73]{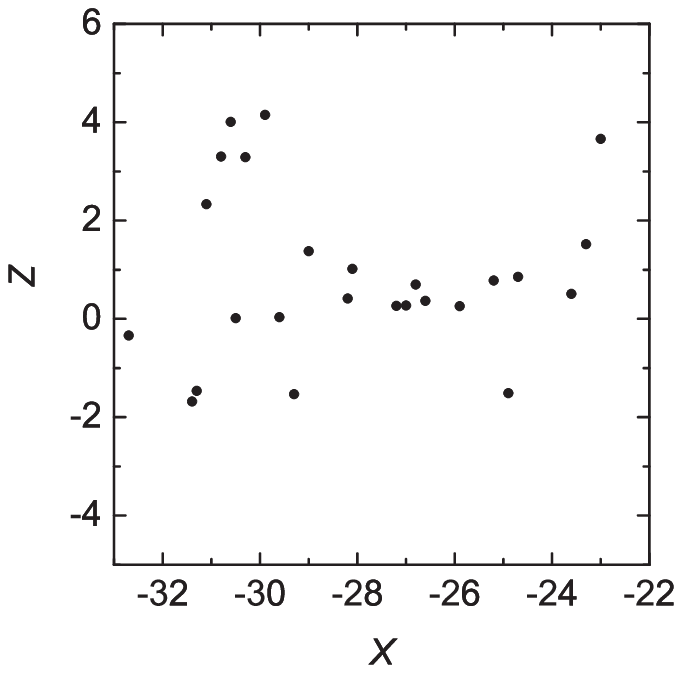}
\hspace{5mm}
\includegraphics[scale=0.73]{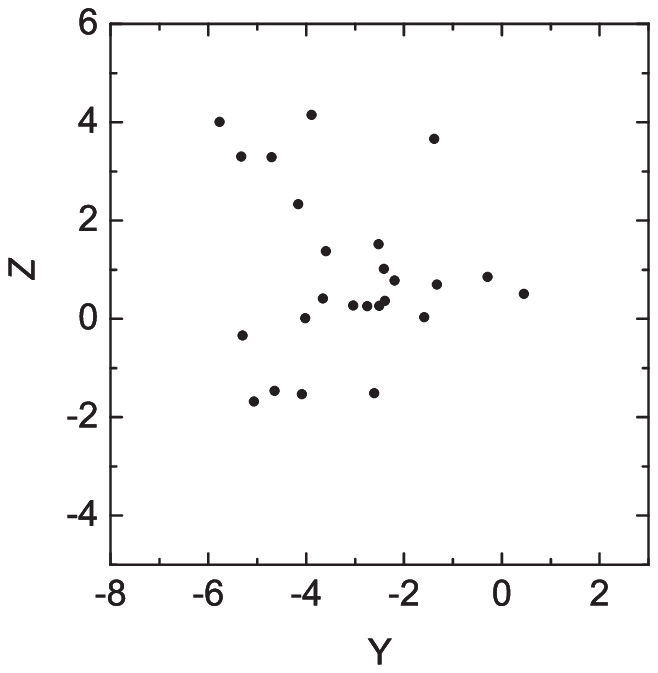}
} \vspace{5mm} \hbox{
\includegraphics[scale=0.73]{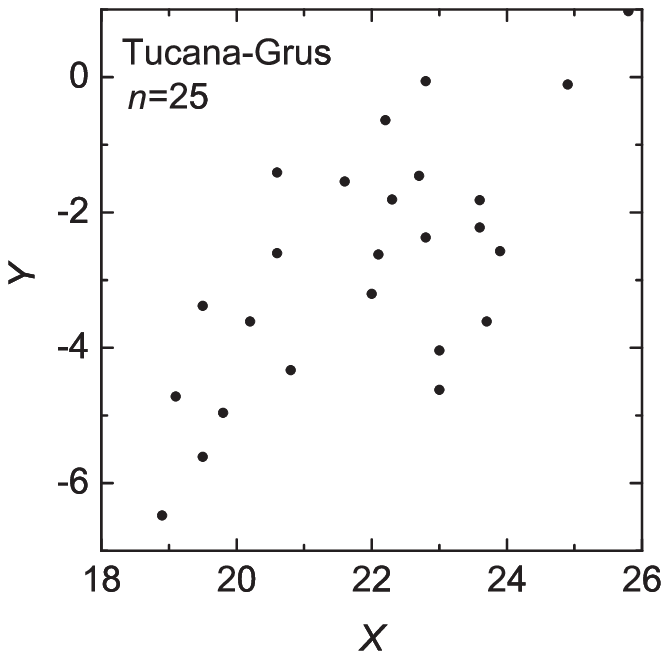}
\hspace{5mm}
\includegraphics[scale=0.73]{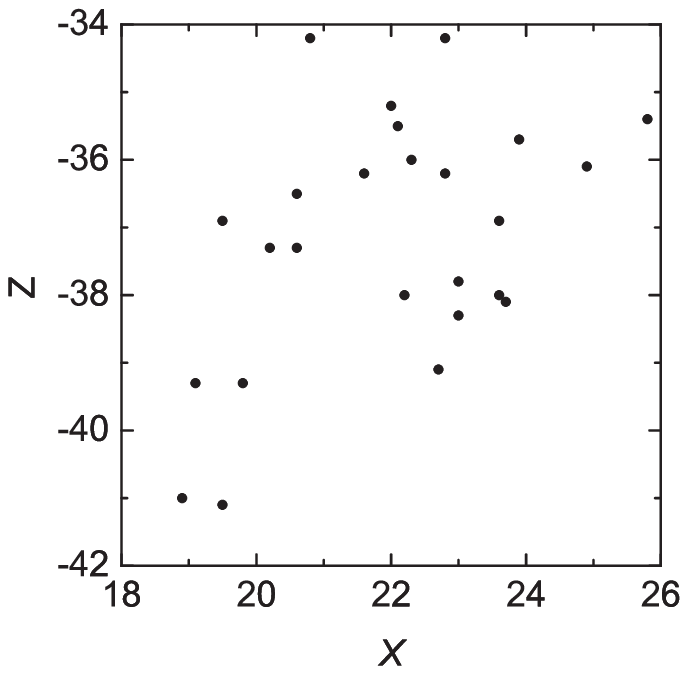}
\hspace{5mm}
\includegraphics[scale=0.73]{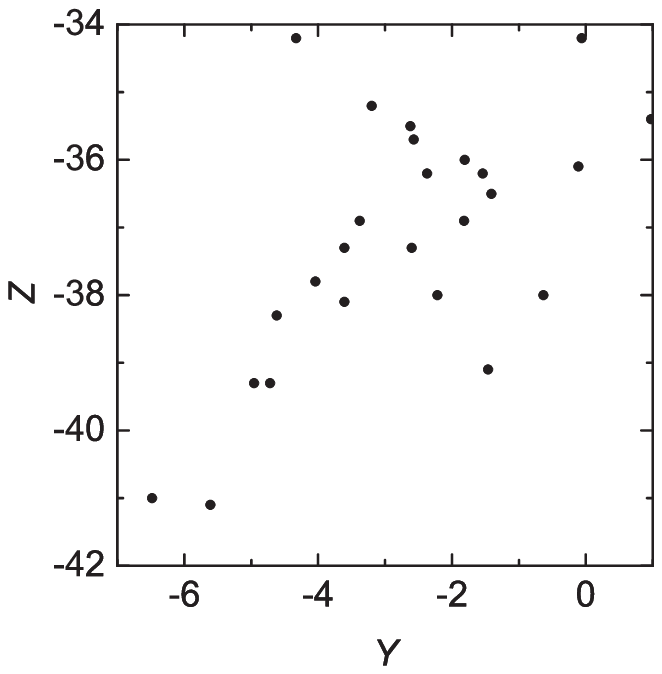}
}}}
\captionstyle{normal} \caption{The structure of eight most
populated low density agglomerates in the projections of
equatorial Cartesian coordinates (in Mpc). The second part,
continued from Fig.~10.}
\end{figure*}

The panels of Figs.~10 and 11 show the distribution of galaxies of
eight most populated agglomerates in the projections of Cartesian
equatorial coordinates (in Mpc). As follows from these figures,
the shapes of agglomerates are very diverse and are generally far
from being spherically symmetric. In a half of cases, they could
be conditionally called filamentary or deplanate.

The median line-of-sight velocity dispersion in rich agglomerates
($170$ km/s)  and median $K$-lumino- sity $(3.0\times
10^{11}~L_{\odot}$) are close to the corresponding median values
for the Makarov-Karachentsev (MK) groups in the same volume of
Local universe~\cite{Mak2011:Karachentsev_n}, but the linear
dimensions of the agglomerates exceed the typical size of
MK-groups by an order of magnitude. The median of the formal value
of virial mass in rich agglomerates, $1.6\times10^{14}~M_{\odot}$
is comparable to the mass of poor clusters, whereas the median of
the formal virial mass-to-$K$-luminosity ratio, equal to about
$700~M_{\odot}/L_{\odot}$ is by an order of magnitude higher than
the corresponding value for the richest clusters.

It should be emphasized that the considered agglomerates are
extremely  incoherent buildups with no obvious signs of
concentration of galaxies towards their geometric centers. The
average number density of galaxies in them is only about five
times higher than the mean number density of galaxies in the
considered volume of the Local universe. Given the scales and
line-of-sight velocity dispersions specified in the table, the
characteristic crossing time in these aggregates is 30--40~Gyr,
what is significantly larger than the age of the Universe.

\section{CONCLUDING REMARKS}

The considered volume of the Local universe with the diameter of
about 100~Mpc is a quite representative sample, including the
entire Local Supercluster and the spurs of other neighboring
superclusters. This volume comprises both groups and clusters, as
well as the cosmic voids. Searching for the diffuse associations
of galaxies in the regions of low density, we used only the
kinematic distances of galaxies,  \mbox{$D=V_{\rm LG}/H_0$},
neglecting their peculiar velocities.  Until fairly recently,
there was an idea that large peculiar velocities of galaxies are
only found  in the ``hot'' virial regions of clusters, while in
the field galaxies the deviations from the Hubble relation
\mbox{$V=H_0D$} are small. According to the current data, however,
the field galaxies, surrounding the Local Group are involved in a
bulk motion towards the Virgo cluster with the velocity of about
$180$~km/s, and in the motion from the center of the Local Void,
caused by the expansion of the void with the velocity of about
$260$~km/s~\cite{Tul2008:Karachentsev_n}. Numerous simulations of
the evolution of large-scale
structure~\cite{Klyp2003:Karachentsev_n,
Schaap2007:Karachentsev_n} reveal the presence of coherent motions
of  field galaxies  with  amplitudes of a few \mbox{hundred km/s}
on the scale of approximately \mbox{($10$--$50$)~Mpc}. The
observational data on large peculiar motions of galaxies in the
Coma I region give indications that there exists a possible ``dark
attractor'' with a mass of around \mbox{$10^{14}~M_{\odot}$} at
the distance of 15~Mpc from us~\cite{Karach2011:Karachentsev_n}.

Large-scale flows of galaxies related to the motions of filaments
and walls can lead to phantom phase groupings of galaxies, if only
the kinematic distances are used for their clustering. Such false
``phase caustics'' can be easily confused with the scattered
physical groupings of galaxies. Therefore, some or even most of
the discussed agglomerates in the low-density regions can prove to
be phantom structures.

It is obvious that for checking the verity of the existence of the
diffuse agglomerates we have discovered, the measurements of
galaxy distances by the Tully-Fisher method
\cite{Tul1977:Karachentsev_n} or any other technique, independent
of the line-of-sight velocities are yet required.


\begin{acknowledgments}
This study was made owing to the support of the following grants:
the grants of the Russian Foundation for Basic Research (project
nos. \linebreak 11-02-90449-Ukr-f-a, 12-02-91338-NNIO), the
Ukrainian State Fund for Fundamental Research (project no.
F40.2/049), the Cosmomicrophysics program of the National Academy
of Sciences of the Ukraine, and by the Ministry Education and
Science of the Russian Federation (state contract
no.~14.740.11.0901).
\end{acknowledgments}

{}

\end{document}